\documentclass[graybox]{svmult}

\usepackage{mathptmx}       
\usepackage{helvet}         
\usepackage{courier}        
\usepackage{type1cm}        
\usepackage{makeidx}         
\usepackage{graphicx}        
\usepackage{multicol}        
\usepackage[bottom]{footmisc}
\usepackage{natbib}

\makeindex             

\begin{document}

\title*{Stellar populations of bulges at low redshift}
\author{Patricia S\'anchez-Bl\'azquez}
\institute{Patricia S\'anchez-Bl\'azquez \at Departamento de F\'{\i}sica Te\'orica, Universidad Aut\'onoma de Madrid, 
\email{p.sanchezblazquez@uam.es}}%
\maketitle

\abstract{This chapter summarizes our current understanding of the 
stellar population properties of bulges and outlines important
future research directions.}

\section{Introduction} 
\label{sec:introduction} 
The stellar populations of bulges 
provide a fossil record of their formation and evolutionary history, including 
insights into the duration and efficiency of the primary epochs of star 
formation. In previous chapters, we have learnt that there are three main types 
of bulges: the classical, the disky, and the boxy/peanut. All of them show 
different structural and kinematical properties and different formation scenarios 
are proposed to explain them. In these scenarios, the star formation history is 
predicted to be different, e.g., the proposed mechanisms to form classical bulges 
imply rapid and efficient star formation while disky bulges are believed to 
form slowly and at lower redshifts from the inflow of mainly gaseous material to 
the center of the galaxy \citep[see][for 
reviews]{wyse1999,kormendykennicutt2004}.

Boxy/peanut bulges are believed to be parts of bars seen edge-on and have their 
origin in vertical instabilities of the disk. Therefore they are expected to have 
a similar stellar population compared to the inner disk. In principle then we should be 
able to distinguish between different formation scenarios simply by studying the 
different ages, metallicities and abundances ratios of the bulges. The situation, 
however, it is not that straightforward; as discussed in other chapters (see 
e.g. Chapter 2.3) internal processes related with disk instabilities can also 
occur at high redshift and in short timescales. Furthermore, bulges with 
properties resembling pseudobulges can form, not only by internal processes 
related with the presence of non-axisymmetric components, but also by accretion of 
gas and galaxies \citep{guedesetal2013, querejeta2014,elichemoral2011, 
obrejaetal2013}. However, getting information on the ages, metallicities and 
abundances ratios in the different types of bulges constitute, undoubtedly, a 
strong constrain for scenarios of bulge formation and, therefore, several authors 
have studied the problem. With the exception of the MW and M31, in which we can 
resolve individual stars, studies of bulges have to deal with integrated 
properties, through their mean color or absorption lines. Such unresolved stellar 
populations studies have been far less common for bulges than for elliptical 
galaxies. The reason is that disk galaxies have more dust and ionized gas. The 
first affects the colors and the second fills the Balmer lines, the most 
important age diagnostics in the optical. In addition, bulges have, in general, 
lower surface brightness than ellipticals and the presence of several 
morphological components, such as disks, bars, rings, etc., complicates the 
interpretation of the results. Lastly, the light coming from the disk may 
contaminate the bulge spectrum in a way that is difficult to quantify. This 
problem is especially acute for studies of stellar population gradients.

Furthermore, over many years, unresolved stellar populations studies have been 
done comparing the integrated colors or absorption lines with the theoretical 
predictions for single stellar populations (SSP); that is, an essentially coeval 
population of stars formed with a given initial mass function with the same 
chemical abundance pattern. While this scenario may not be a bad approximation 
for massive elliptical galaxies, bulges, especially those formed secularly, are 
believed to have a more extended star formation history. This means that the young 
populations, which have low mass-to-light ratios, bias the analyses of 
composite populations, if present \citep[e.g.][]{trageretal2000}.

The relatively low number of studies, the small -- and biased -- samples, and the 
difficulties pointed out above have led to a lack of consensus about important 
results concerning the stellar populations of bulges, as I will show in this 
review. However, in the last decade, stellar population models which predict not 
only individual spectral features, but the entire synthetic spectra for a 
population of a given age and metallicity \citep{vazdekis1999, 
bruzualycharlot2003, vazdekisetal2010, coehloetal2007, walcheretal2009, 
conroyetal2014} have been released. The availability of these models is 
stimulating the development of numerical algorithms to invert the observed galaxy 
spectrum onto a basis of independent components (combination of single stellar 
populations, age-metallicity relation, and dust extinction).  Also, new 
specialized software allow the separation of the light coming from the stars and 
ionized gas in a reliable way \citep[e.g.][]{sarzietal2006}. In addition to this, 
new data from integral field spectrographs \citep[e.g.][]{baconetal2001, 
cappelari2011, blancetal2013} are changing the way we see galaxies 
\citep{sanchezcalifa, rosalesortega2010}. The analysis of these datasets allows 
one to associate stellar population properties with morphological and kinematical 
characteristics of the galaxies, making the interpretation of stellar populations 
more secure. Therefore, the development of the field is very promising and we 
foresee important advances in the decades to come.

In this chapter, I will try to review the state of the art in the area, trying to 
highlight the necessary steps to get a better understanding of the star formation 
histories of these complex systems. Section~\ref{sec:colors} summarizes the 
general results obtained with single apertures. Section~\ref{sec:bars} compile 
the works on the possible influence of bars in the stellar populations of bulges 
and on the stellar populations of bars themselves. Section~\ref{sec:sfh} outline 
the results obtained with full spectral fitting techniques and 
Sect.~\ref{sec:gradients} reports on the studies of stellar population gradients 
in bulges. In Sect.~\ref{sec:connection}, I show the results about the possible 
connection between the stellar populations of bulges and disks while in 
Sect.~\ref{sec:summary} the main results are summarized. In Sect.~\ref{sec:future}, 
I give some thoughts of what I think the next steps for the study of stellar 
populations in bulges should be.

\section{Results obtained with single aperture}
\label{sec:colors}

\subsection{General properties}

The first studies of stellar populations in bulges were performed using optical 
and near-infrared broadband photometry \citep{peletierbalcells1994, 
terndrupetal1994, peletierbalcells1996, belldejong2000, dejong1996}. These works 
demonstrated that changes in the bulge colors are linked to galaxy luminosity, 
potential well, and local surface brightness, with more massive/luminous bulges 
and higher surface brightness regions being redder than less massive/luminous and 
lower surface brightness ones. They also showed that early-type bulges are red, 
as red as elliptical galaxies, and with very little dispersion in their colors 
\citep{peletierbalcells1996}. These results do not apply to the few late-type 
galaxies analyzed, where significantly bluer colors are measured.

Early interpretation of these data pointed to early-type bulges being as old as 
ellipticals with late-type, less massive ones being younger and/or more metal 
poor. The small dispersion in the colors was interpreted as being due to small 
dispersion in the age of early-type bulges (maximum of $\sim$2 Gyr). This, in 
principle, is in agreement with the classical and disky bulge formation scenarios 
(see Sect.~\ref{sec:introduction}) if, as it seems the case, secularly formed 
disky bulges are more common in late-type and in less massive galaxies 
\citep{kormendykennicutt2004, gandaetal2009}\footnote{Note, however, that a 
significant number of local massive spiral galaxies appears to have dominant 
pseudobulges, defined as those bulges with $n<2$, that includes both disky and 
boxy/peanut bulges \citep{kormendyetal2010}. Furthermore, pseudobulges are also found in S0 galaxies (Laurikainen et al. 2012).}.

One problem of using colors is the well known age-metallicity degeneracy 
\citep{worthey1994}. Bluer bulges can be either younger or more metal poor and, 
without this information, it is difficult to extract conclusions about their 
formation mechanisms. Further complications are the presence of emission lines 
and dust extinction that also affect the colors. In particular, dust extinction 
depends on the inclination and, therefore, inclination is another parameter that 
needs to be taken into account when comparing the colors of different types of 
bulges and also when comparing the results from different studies 
\citep[see][]{gandaetal2009}. All these obstacles make very difficult to extract 
useful conclusions about the stellar populations of bulges using only colors. For 
these reasons studies with colors need to be complemented with those using 
information of the absorption lines with different sensitivities to age and 
metallicity and that are not affected by dust extinction \citep{macarthur2005}.

The first spectroscopic studies of bulges analyzed the relation of line-strength 
indices \citep[the so-called Lick/IDS indices, see][]{gorgasetal1993, 
wortheyetal1994} with the central velocity dispersion ($\sigma$ hereafter) -- 
used as a proxy for the dynamical mass of the galaxy. Lick/IDS indices measure 
the strength of the most prominent absorption lines in the optical galaxy spectra 
and are sensitive to changes of the mean age, chemical abundances and, to a 
lesser extent, the initial mass function \citep[e.g.][]{vazdekisetal2010, 
bruzualycharlot2003, thomas2003, schiavon2007, conroyvandokkum2012}. These 
studies confirmed the similarity of bulges with elliptical galaxies for 
early-type galaxies \citep[earlier than Sbc,][]{benderetal1993, fisheretal1996, 
idiartetal1996}.

However, it has been pointed out that this similarity may be due to the fact that 
the majority of these early analyses were performed on samples that are biased 
towards early-type spirals \citep[earlier than Sbc,][]{kormendykennicutt2004}. In 
the last few years, however, several studies have included in their samples 
late-type bulges and analyzed, mostly, the relation between the Mg-sensitive 
indices (Mg$_2$ and Mgb) and the central $\sigma$. When these bulges were included, 
differences between elliptical and bulges were found. However, the nature of 
these differences is still not clear. Some authors claim that bulges are located 
below the Mg-$\sigma$ relation obtained for ellipticals, which is commonly 
interpreted as bulges having a younger stellar population 
\citep{prugnieletal2001,chiappinietal2001,gandaetal2007, morellietal2008}. Other 
authors find that the slope of the Mg-$\sigma$ is steeper for bulges 
\citep{falconbarrosoetal2002}, while \citet{trageretal1999} and 
\citet{proctorsansom2002} report that only low-mass bulges depart from the 
relation between spectral indices and $\sigma$ drawn by large bulges. On the 
other hand, other studies do not find any systematic difference in the 
Mg-$\sigma$ relation of bulges and elliptical galaxies but find that the scatter 
among this relation for bulges is larger than the equivalent one for ellipticals 
\citep{moorthyholtzman2006,peletieretal2007}. Similar conclusions were obtained 
using other line-strength indices.

\begin{figure}
\centering
\includegraphics[width=\hsize]{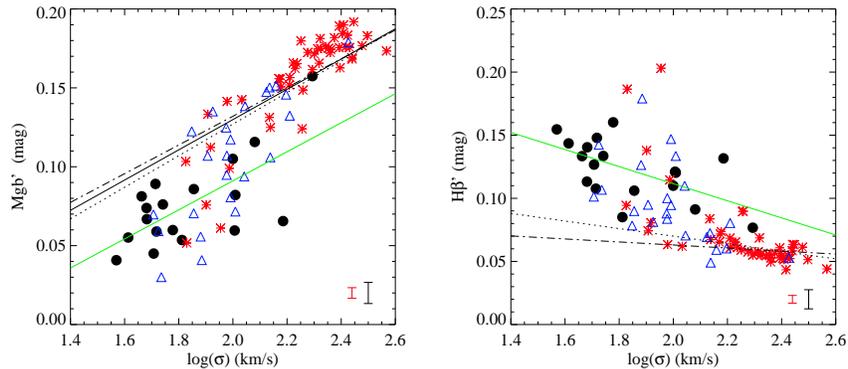}
\caption{Line-strength indices Mgb$'$ and H$\beta '$ expressed in 
magnitudes, against central velocity dispersion. The black symbols 
represent the sample of late-type bulges from \citet{gandaetal2007}, and the 
blue and the red symbols, the sample of Sa galaxies and E and S0 
respectively from the SAURON survey. The dotted and dashed-dotted black lines over-plotted in 
both panels are the relations obtained by \citet{sanchezblazquez2006a} 
for low- and high-density environments, respectively, and the green 
solid lines are the relations determined using the late-type galaxy 
sample. Representative error bars are added at the bottom right of each 
panel; the black one refers to both the early- and late-type spiral 
samples, while the red one refers to the E/S0 galaxies. Figure taken 
from \citet{gandaetal2007}.\label{fig:ganda2007}}
\end{figure}

Some of the discrepancies in the conclusions of different studies may be due to 
differences in the mass distribution (or central $\sigma$) of the selected 
sample. For example, Fig.~\ref{fig:ganda2007} shows the relation between the Mgb 
and H$\beta$ indices (measured in magnitudes) for a sample of late-type and 
early-type bulges and elliptical galaxies from \citet{gandaetal2007}. It can be 
seen that despite the relations between line-strength and indices followed by 
late-type bulges and early-type galaxies running apparently parallel to each other, the 
differences may just be related with the different range of central $\sigma$. In 
fact, the differences disappear at low-$\sigma$, where even early-type galaxies 
(E and S0\footnote{Classically, studies of stellar populations include E and S0 
galaxies in the same group.}) deviate from the relation defined by massive 
ellipticals.

Furthermore, bulges, contrary to massive elliptical galaxies, are rotationally 
supported. Some authors have cautioned 
\citep{prugnielsimien1994,falconbarrosoetal2002} that by not taking into account 
the rotation in the $\sigma$ measurements, one may be underestimating their 
binding energy. The contribution to the rotation may by calculated as $0.5 
\log(1+0.62 V^2/\sigma^2)$, as in \citet{prugnielsimien1994}, where $V$ is the 
rotational velocity. \citet{falconbarrosoetal2002} claim that a mean $V/\sigma = 
0.5$ suffices to bring the bulges back to the Mg$_2$-$\sigma$ relation defined by 
giant ellipticals\footnote{Note, however, that the majority of low-luminosity 
elliptical galaxies are also rotationally supported \citep{emsellemetal2011}.}. 
Other studies have also claimed a better correlation between the line-strength 
indices and $V_{\rm max}$ (an indicator of the total potential well) than between 
line-strength indices and the central $\sigma$ \citep{prugnieletal2001}.

\subsection{Comparison with SSP models} 

\begin{figure} 
\centering 
\includegraphics[scale=.5]{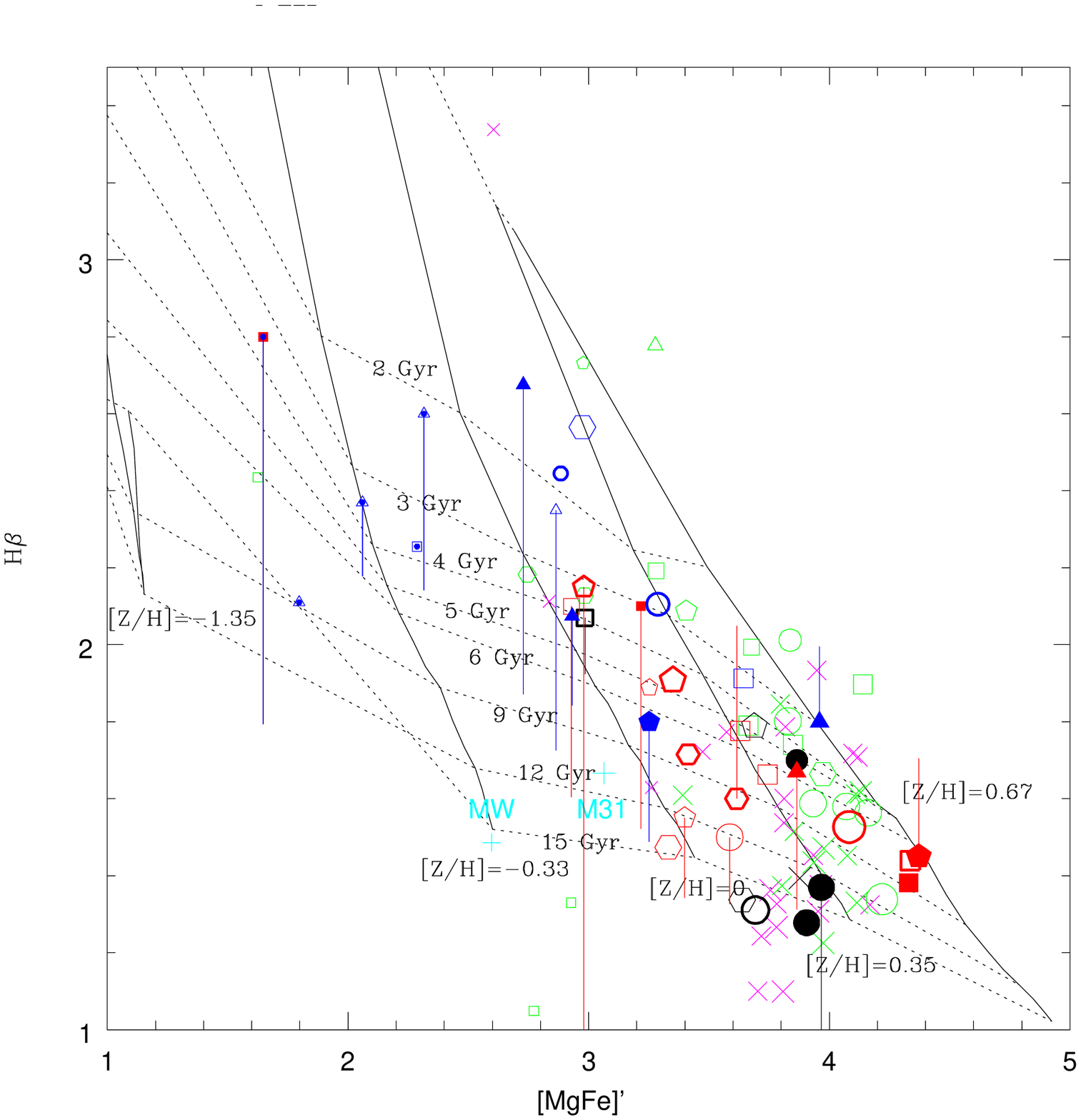} 
\caption{H$\beta$ vs. $[$MgFe$]'$ in the central regions of bulges and 
ellipticals. Magenta crosses are galaxies from \citet{trageretal1999}. 
Green symbols are bulges and ellipticals from \citet{proctorsansom2002}. 
The MW \citep{puziaetal2002} and M31 \citep{puziaetal2005} are shown as '+' 
symbols. Blue and red symbols are from \citet{moorthyholtzman2006}. The 
color of the symbol is chosen according to whether they are redder or 
bluer than B-K=4. Larger sizes indicate larger central velocity 
dispersion. Bulges for which there is no color information are in black. 
Solid lines represent the predictions of \citet{thomas2003} for SSPs of 
constant metallicity (as indicated in the labels) while dotted lines 
represent the predictions for populations of constant age, with age 
increasing towards the bottom of the panel. For other details regarding 
the figure see \citet{moorthyholtzman2006}.
\label{index_index}}
\end{figure}

Nevertheless, the similarity found by 
some authors between the index-$\sigma$ relation of bulges and elliptical 
galaxies may not reflect a real similarity in their stellar content. 
Line-strength indices are not free from the age-metallicity degeneracy (e.g., 
Mg$_2$ and Mgb can be lower in younger or in more metal stellar populations). 
Therefore, large differences between the ages of bulges and ellipticals could 
exist and not be reflected in these relations if there is a complementary 
age-metallicity relation \citep[e.g.][]{trageretal2000}. The advantage of using 
these characteristics, though, is that the sensitivity to variations of age and 
metallicity of each different index varies. A way to partially break the 
age-metallicity degeneracy is to combine indices more sensitive to mean age 
variations (i.e., the Balmer lines) with those more sensitive to abundance 
variations in the so-called index-index diagrams. Figure~\ref{index_index} shows 
one of these diagrams combining the composite index \[ 
[\textrm{MgFe}]'=\sqrt{\textrm{Mgb} (0.28 \textrm{Fe5270}+ 0.72 
\textrm{Fe5335})},\] which is fairly insensitive to variations of $[\alpha/$Fe$]$ 
abundances\footnote{The $\alpha$ elements are those chemical elements 
predominantly formed via fusion with a helium nucleus. Their most abundant 
isotopes therefore have nucleon numbers that are multiples of four (e.g., O, Ne, 
Mg, Si, S, Ar, Ca, Ti). These elements are mainly synthesized in Type II 
supernovae, while Type Ia supernovae produce elements of the iron peak (V, Cr, 
Mn, Fe, Co and Ni). In chemical evolution models, type II supernovae produce an 
early enrichment of $\alpha$-elements followed by a subsequent enrichment of 
iron-peak, Type Ia supernovae products. In the absence of other modifying 
factors, this implies that $[\alpha$/Fe$]$ can be used as a 'galactic clock' for 
the duration of the star formation.} \citep{thomas2003} and the Balmer index 
H$\beta$.

Several authors have used this technique to compare the index values with the 
predictions of SSP models. These comparisons show that bulges have a large range 
in SSP-equivalent ages from $\sim$2 to 13.5 Gyr 
\citep{peletieretal2007,moorthyholtzman2006} and metallicities. They also report 
a correlation between both the SSP-equivalent age and metallicity and central 
$\sigma$. In general, they found that more luminous/massive bulges were older and 
more metal rich. They also inferred that more massive bulges have a larger ratio 
of $\alpha$-elements\footnote{What it is usually measured is the Mg abundance 
through the Mgb index. Other $\alpha$-elements, like Ca or Ti may follow 
different patterns 
\citep[e.g.][]{conroyetal2014,gravesschiavon2008,cenarroetal2004}.} with respect 
to Fe, which is usually interpreted as more massive bulges forming their stars on 
shorter timescales. The relations were similar to those found for elliptical 
galaxies 
\citep{bica1988,jablonkaetal1996,idiartetal1996,casusoetal1996,goudfrooijetal1999, 
trageretal1999,thomasdavies2006,moorthyholtzman2006,jablonkaetal2007,gandaetal2007}. 
On the other hand, \citet{proctorsansom2002} and \citet{prugnieletal2001} found 
that, contrary to what happens in elliptical galaxies, both Fe and Mg were 
correlated with $\sigma$ in bulges, resulting in the lack of a tight correlation 
between Mg/Fe (a proxy for $[\alpha/{\rm Fe}]$) and $\sigma$. This result needs to be 
corroborated by other studies.

\begin{figure}
\includegraphics[trim=20 0 20 0,clip,scale=.62]{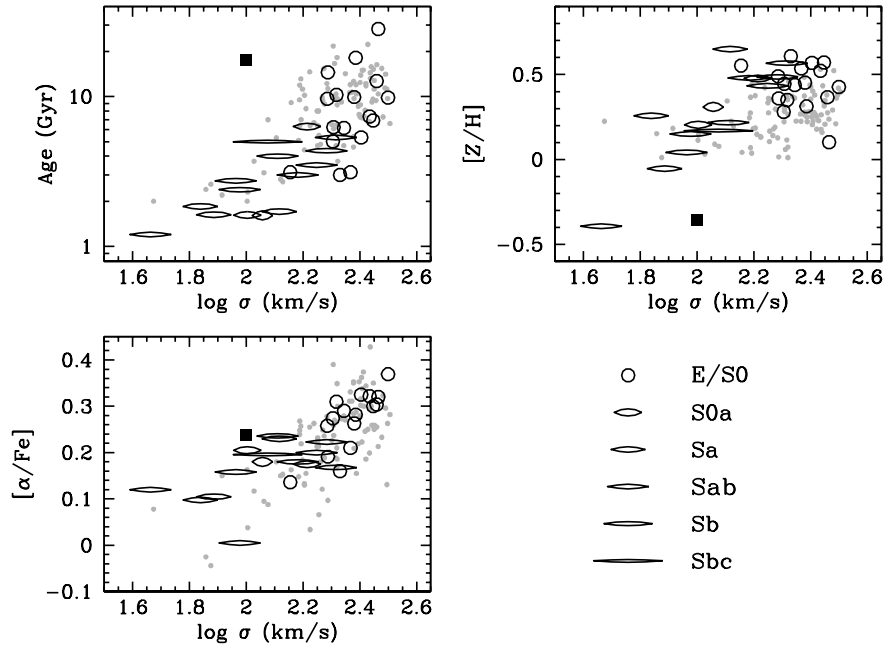}
\caption{Stellar population parameters {\it versus} central velocity 
dispersion. Open circles are early-type galaxies from  
\citet{thomasdavies2006}, ellipses are spiral bulges with ellipticity 
increasing for the later-types (see labels in the right-hand bottom 
panel) and the filled square is the integrated light of the MW bulge. 
Small grey-filled circles are early-type galaxies from 
\citet{thomasetal2005}. Central stellar populations are shown.
Figure taken from \citet{thomasdavies2006}.
\label{thomasdavies}}
\end{figure}

Similar to the results obtained with line-strength indices, the comparison of 
SSP-equivalent parameters of bulges and ellipticals reveal that both have very 
similar properties, at least in samples of bulges earlier than Sbc. 
Figure~\ref{thomasdavies} illustrates the relation between the SSP-equivalent 
ages, metallicities, and $[\alpha/{\rm Fe}]$, and the central $\sigma$ for bulges, S0, 
and elliptical galaxies \citep{thomasdavies2006}. It can be seen that, at a given 
central $\sigma$, the stellar population parameters of bulges and elliptical 
galaxies are indistinguishable. These results support the idea that bulges (with 
morphological types earlier than Sbc) were formed with very little influence from 
the disk, in a process similar to the one that formed elliptical galaxies.

In samples that contained late-type galaxies (later than Sbc), both 
\citet{prugnieletal2001} and \citet[][see also \citealt{morellietal2008}]{moorthyholtzman2006} find three types of bulges in the comparison of 
line-strength indices and stellar population models: the old-metal rich (OMR), 
the young metal rich (YMR) -- which are bulges with ages less than 3~Gyr and 
super-solar metallicities -- and the metal poor (MP), with sub-solar metallicity. 
These classes seem to be sensitive to the Hubble type. All the early-type 
(S0-Sab) bulges are metal-rich. The red early-type bulges are in the OMR region 
while the blue early-types reside in the YMR region. Metal-poor bulges are all 
late-types, but late-type bulges are found in all three regions. A comparison of 
the SSP properties of late-type bulges and ellipticals galaxies at a similar 
$\sigma$, however, remains to be done and, therefore, as it was the case with the 
line-strength indices, it is not clear if late-type bulges are younger, for being 
late-type, or for having low $\sigma$.

\subsection{Relation between stellar population and structural properties}

\begin{figure} \centering \includegraphics[scale=.65]{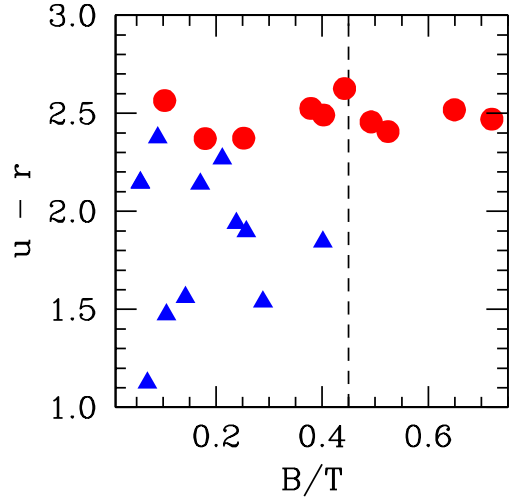} 
\caption{Distribution of bulge-to-total ratios, B/T, of intermediate type 
(Sa-Sbc) galaxies with pseudobulges (blue triangles) and classical bulges (red 
circles) with respect to their global u-r color. The dashed line marks B/T=0.45 
above which only classical bulges are found. From \citet{droryfisher2007}. 
\label{fisher_drory}} \end{figure}

A more direct way to test the different proposed scenarios for bulge formation is 
to compare the stellar population properties of a sample of bulges with 
morphological or dynamical properties that distinguish them as classical, disky, 
and boxy/peanut. As we have seen in previous chapters (see Chapter 2.2), 
there are several observables used to separate bulges and pseudobulges and 
different authors have employed different properties to perform this task. 
\citet{carolloetal2001} analyzed V, H, and J HST images of a sample of bulges 
with exponential (typical of disky and boxy/peanut bulges) and R$^{1/4}$ 
luminosity profiles (typical of classical bulges), finding the former, on 
average, bluer than the latter (by $\sim$0.4 mag in $<V-H>$), which could be the 
consequence of a younger and/or of a lower metallicity stellar population. They 
also found, in agreement with the results of \citet{peletierbalcells1996}, that 
the colors of those bulges showing a R$^{1/4}$ profile were red and very 
homogeneous, while for the exponential bulges the scatter was significantly 
larger\footnote{Note, however, that many of the exponential bulges were showing 
colors as red as the R$^{1/4}$ bulges.}. They interpreted these results as a 
delayed formation of the exponential bulges compared with those having an 
R$^{1/4}$ profile, which formed their stars in the early Universe. 
\citet{droryfisher2007} used a different approach and separated classical and 
pseudobulges morphologically. Pseudobulges were those showing nuclear bars, 
nuclear spirals, and/or nuclear rings and classical bulges those featureless 
structures rounder than the outer disk. Separating the bulges in this way and 
comparing with their visual morphological types, they studied the location of 
bulges in the color-magnitude diagram, finding that Sc galaxies and later types 
do not contain classical bulges and are located almost entirely in the blue cloud 
in the color-magnitude diagram. Intermediate Sa-Sbc type galaxies, on the 
contrary, contain both classical and pseudobulges. While 87\% of the galaxies 
with pseudobulges were in the blue cloud, all galaxies with classical bulges were 
in the red-sequence. These authors stress that the differences in colors are not 
due to a different contribution of bulge and disk to the total galaxy color as 
blue and red galaxies share a range in bulge-to-total ratio. This is shown in 
Fig.~\ref{fisher_drory}, where the global color of the galaxies is plotted as a 
function of the bulge-to-total (B/T) ratio. It can be seen that both bulges and 
pseudobulges coexist in the region of B/T values ranging from 0.05 to 0.45 and 
that, in this region, classical bulges are redder than pseudobulges. They also 
found that, in general, pseudobulges classified with the morphological features 
described above were more diffuse and had lower S\'ersic indices than classical 
bulges. 

Another way of separating classical bulges from pseudobulges was adopted by 
\citet{gadotti2009} who identified pseudobulges as those lying below the Kormendy 
relation \citep{kormendy1977} defined by elliptical galaxies. He established that 
pseudobulges defined this way were, in general, 0.2~mag bluer in the ($g-i$) 
color than the classical bulges. On the other hand, 
\citet{fernandezlorenzoetal2014}, who used the S\'ersic index to differentiate 
between classical and pseudobulges, found the latter as red as the former at the 
same luminosity. Only in fainter pseudobulges they did measure bluer colors.

Again, colors are affected by the age-metallicity degeneracy and dust extinction and, in 
principle, more information may be obtained from the study of absorption lines 
and their comparison with stellar population models. Several works have also 
compared the line-strength indices of bulges with different structural 
characteristics \citep{williamsetal2012, gadotti2009}. The first studied a sample 
of edge-on boxy/peanut bulges finding that they follow the same central 
index-$\sigma$ relation as elliptical galaxies \citep[see 
also][]{jablonkaetal2007}, although the sample was biased towards early-type 
galaxies (S0-Sb). On the other hand, \citet{gadotti2009} found a strong 
correlation between the D4000 index and the S\'ersic index ($n$), indicating younger 
populations\footnote{Actually the author did not compare the index with stellar 
population models and, therefore, a variation in metallicity is also possible.} 
in galaxies with lower $n$. On the contrary, for a sample of early-type galaxies, 
\citet{vazdekisetal2004} did not find a correlation between $n$ and age, while 
they found a strong correlation between $n$ and $[$Mg/Fe$]$. The different 
behavior for bulges and ellipticals is very interesting, but more studies of this 
kind using larger samples are still needed to confirm or refute the trends. Note 
that the only study targeting specifically boxy/peanut bulges is that of 
\citet{williamsetal2012}. The rest of studies cited above usually include both 
disky and boxy/peanut bulges in the same category, the `pseudobulge' sample.

The current lack of consensus between studies may be due to different criteria 
to separate classical and pseudobulges. Some might include disky and 
boxy/peanut bulges in the same category, pseudobulge, without making 
any distinction between them. Furthermore, different distributions of 
the galaxy luminosities can also lead to discordant results. It is clear 
that low-luminosity, low-mass bulges are bluer than more massive and 
brighter ones, but it is not clear if, at the same luminosity, bulges 
with different structural characteristics share the same color.

\subsection{Bulges as composite systems}
\label{sec:composite}

Thanks to the 2-dimensional data of the SAURON survey \citep{baconetal2001}, 
\citet{peletieretal2007} \citep[see also][]{silchenkoafanasiev2004} noticed that, 
when present, young stellar populations in their sample of early-type bulges were 
concentrated near the center, in disks or in annuli suggestive of resonance rings 
\citep{byrdetal1994}. \citet{peletieretal2007} realized that the studies 
comparing the line-strength indices of bulges and elliptical galaxies could be 
divided into two categories: those targeting inclined galaxies, which do not find 
any difference between the index-$\sigma$ relation of bulges and ellipticals, and 
those sampling almost face-on galaxies, which find younger stellar populations in 
bulges compared with those of elliptical galaxies and a large scatter in the 
line-strength indices at a given $\sigma$. The differences are especially visible 
in galaxies with low $\sigma$.  They argue that bulges are composite systems, 
with two or more types of bulges coexisting in the same galaxies. The classical 
bulge is composed mainly of an old and metal rich population and the disky and 
boxy/peanut bulge can be younger and contain more metal-poor stars (although it 
can also be old). The discrepant results obtained in samples of different 
inclinations can be explained, according to these authors, by the different 
contribution to the bulge light of different subcomponents (classical, disky, and 
boxy/peanut). If the young component is a disk, then it is concentrated in a 
plane and it would not be observed in edge-on galaxies. These young components, however, do 
contribute to the integrated light of less inclined samples (if they are limited 
to the central regions). The result is supported by the observation of central 
dips in the velocity dispersion maps in 50\% of the galaxies of their sample.

The coexistence of 2 or more types of bulges (classical, disky, and boxy/peanut) 
in some galaxies has been pointed out by several authors 
\citep{athanassoula2005,gadotti2009,nowaketal2010, kormendybarentine2010, 
erwin2008} and is supported by theoretical studies 
\citep{obrejaetal2013,samlandgerhard2003}. \citet{obrejaetal2013} propose a 
picture were the centers of most early-type spirals contain multiple kinematic 
components: an old and slowly rotating elliptical-like component, and one or more 
disk-like, rotationally supported components which are typically young but can 
also be old.

This `two component model' also explains the properties of our MW bulge. The 
MW is considered to have a boxy bulge, yet increased evidence of an old, 
$\alpha$-enriched stellar population that formed on a short time-scale has 
resulted in a two component model \citep[e.g.][]{tsujimotobejji2012}. It has been 
shown that two stellar populations coexist in the Bulge separated in age and 
metallicity \citep{mcwilliamrich1994, feltzinggilmore2000, vanLoon2003, 
groenewegen2005, zoccalietal2006, fulbright2007, zoccalietal2008} and that the 
separation extends somewhat to kinematics \citep{zhaoetal1994, sotoetal2007}, 
even if age determinations through color-magnitude diagram shows that most bulge 
stars in the Galaxy are older than 10 Gyr \citep{ortolanietal1995, 
feltzinggilmore2000, zoccalietal2006, clarkson2008}.

The task of isolating the stellar population properties of the different 
subcomponents forming a bulge is difficult. Still, it has been tried by some 
authors. For example, \citet{williametal2011} study the stellar populations of two 
edge-on boxy/peanut shaped bulges. They place the slit along the major axis and 
observe with three offset in parallel positions. They found that NGC~1381 has a 
boxy bulge, with stellar rotation neither cylindrical (as would be expected for 
bars seen edge-on) nor strongly non-cylindrical and with a double hump on the 
rotation curve. The galaxy shows a metallicity gradient but no age gradient and a 
positive $[\alpha/$Fe$]$. They explain the properties of these galaxies in an 
scenario where NGC~1381 has the three classes of bulges. The classical bulge 
formed their stars rapidly and explain the general trend in $[\alpha/$Fe$]$ as a 
function of height, as disk light (with its lower $[\alpha/$Fe$]$) contributes 
less and less to the integrated spectrum. The boxy appearance is explained by the 
simultaneous presence of a bar (which appears boxy in projection), and the double 
hump of the rotation curve hints at the presence of a small disky pseudobulge 
\citep[see also][]{silchenko2010}.

\section{The influence of bars in building up the bulge}
\label{sec:bars}

It seems clear that some bulges have central disks \citep{peletieretal2007}, 
often (but not always) with young stars, which is usually linked to disk gas 
inflow and central star formation caused by internal secular processes related 
with the presence of a bar \citep{friedlibenz1995, normanetal1996, noguchi2000, 
immelietal2004}. However, this central rotationally younger component does not 
necessarily form due to internal processes. Major and minor mergers and external 
accretion of gas may result in the formation of a disky bulge 
\citep[e.g.][]{guedesetal2013, querejeta2014}. This idea may be supported by some 
observational studies. For example, \citet{kannappanetal2004} found, in a sample 
of disky bulges selected to be blue and, therefore, with central young stars, 
that all of them showed signs of recent interactions.

Hydrodynamical cosmological simulations predict that both secular and 
external processes contribute to create disky pseudobulges with 
similar characteristics in both cases, rotationally supported and with  
a young 
and metal poor stellar population \citep{obrejaetal2013, 
guedesetal2013}. \citet{elichemoral2011} also analyze the effects of minor 
mergers on the inner part of disk galaxies, finding this process to be 
efficient in forming rotationally supported stellar inner components, 
i.e., disks, rings or spiral patterns  
\citep[see also][]{domingueztenreiro1998, aguerri2001, scanapiecco2010}.

A way to quantify the importance of secularly formed disky bulges is to compare 
properties of galaxies with and without bars. While it has been found that 
H$_{\alpha}$ emission is enhanced in early-type spirals with bars with respect to 
those early-type non-barred \citep[e.g.][]{hoetal1997, huangetal1996, 
alonsoherreroknapen2001,jogeeetal2005, ellisonetal2011}, the evidence supporting 
the bulge building by bars from the ages of its stars has proven to be elusive. 
Several authors have tried the comparison using samples of face-on galaxies, 
where it is easy to morphologically identify the bar. The differences, however, 
have not been firmly established. \citet{moorthyholtzman2006} and 
\citet{perezsanchezblazquez2011} found hints of lower ages and higher 
metallicities in barred galaxies compared to their counterparts in unbarred at a 
given $\sigma$\footnote{Although the differences found by 
\citet{moorthyholtzman2006} in the H$\beta$-$\sigma$ relation between barred and 
non-barred galaxies disappear when $V_{\rm max}$ is used instead of $\sigma$.}. 
They also found higher $[\alpha/$Fe$]$ abundances in barred galaxies with central 
velocity dispersion 2.2 $> \log \sigma$ (km/s) $ >$ 2.35, but the opposite for $2 > 
\log \sigma$ (km/s) $> 2.2$. At fixed $\sigma$ and $V_{\rm max}$, barred galaxies 
appear to have larger central values of $[$MgFe$]'$\footnote{This index is 
defined as $[$MgFe$]' \equiv \sqrt{\textrm{Mgb}\times (0.72 \textrm{ Fe5270} + 
0.28 \textrm{ Fe5335}}$ in \citet{thomas2003}.} (which can be used as an indicator 
of metallicity independent of $[\alpha/$Fe$]$, see above) than non-barred 
galaxies (or galaxies with elliptical shape bulges) of the same $\sigma$ or 
$V_{\rm max}$. The differences, however, were not very significant in a 
statistical sense.

On the other hand, \citet{jablonkaetal2007} found no difference between the 
stellar population properties of edge-on barred and non-barred galaxies. However, 
it may be difficult to detect a bar in an edge-on galaxy. 
\citet{lorenzocaceres2012} and \citet{lorenzocaceres2013} analyzed the stellar 
populations in the center of double-barred early-type S0s and spirals finding some 
signs of gaseous flows and young stellar populations. This population was not 
very prominent though. Nevertheless, all the above studies were affected by poor 
number statistics.

\begin{figure}
\centering
\includegraphics[scale=.55]{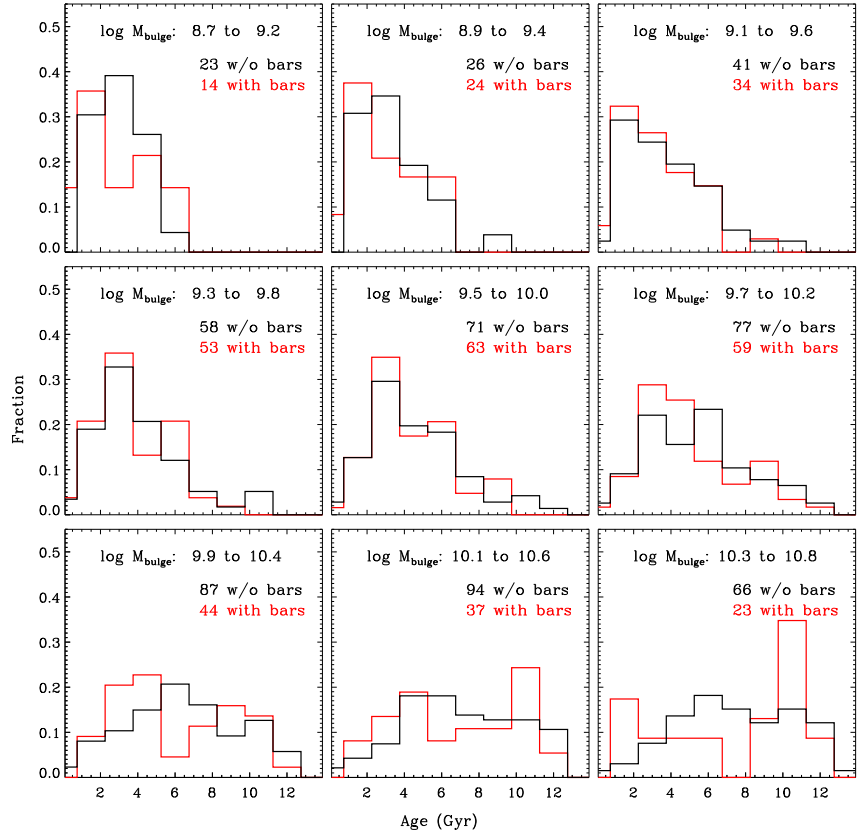}
\caption{Normalized distributions for bulge ages, for several mass 
intervals, as indicated. Distributions for barred and non-barred galaxies 
are shown in red and black lines, respectively. \citep[Figure taken from][]{coehlogadotti2011}.
\label{coleh_gadotti}}
\end{figure}

\citet{coehlogadotti2011} observed 575 face-on bulges in disk galaxies, of which 251
contain bars. They found that, for bulges with masses between 
10$^{10.1}$M$_{\odot}$ and 10$^{10.85}$M$_{\odot}$, the distribution of ages in 
barred galaxies is bimodal with peaks at 4.7 and 10.4~Gyr. This bimodality is not 
seen in non-barred galaxies of a similar bulge mass range. The age distribution 
of barred and non-barred galaxies is, per contra, similar for bulges of masses 
lower than $10^{10.1}$M$_{\odot}$ (i.e., the differences are only seen in massive 
bulges). These authors did not find any difference in the metallicity 
distribution of barred and non-barred galaxies. These results are summarized in 
Fig.~\ref{coleh_gadotti} where the distribution of ages for several mass 
intervals is shown for samples of barred and non-barred galaxies.

Therefore, it is still not clear if there are differences between the stellar 
populations of galaxies with and without bars. What seems clear though is that, 
if there are differences, they are only visible in massive, early-type galaxies, 
a result that is supported by other studies analyzing the molecular gas 
concentration \citep{sakamotoetal1999}. This is often attributed to the fact that 
early-type galaxies host larger and stronger bars than late-type galaxies 
\citep[e.g.][]{butacombes1996}, although, as pointed out in 
\citet{laurikainen2007} and \citet{laurikainen2004}, a longer bar does not 
implies a stronger bar and, in fact, in early-type galaxies the bar induced 
tangential forces are weaker because they are diluted by the more massive 
bulges\footnote{Although the majority of authors does not distinguish different 
types of S0, \citet{laurikainen2007} also show that early-type S0s have shorter 
bars than later type S0s, i.e., the trend of longer bars for early-type 
morphologies reverse in this morphological subclass.}.

This does not mean, necessarily, that bars are not efficient agents in building 
up bulges. It is still not clear if bars are long-lasting structures or not. If 
bars are not long-lasting structures but recurrent patterns \citep{bournau2002} 
then the fact that we do not find differences between barred and non-barred 
galaxies would not necessarily imply that bars are not important for secular 
evolution but, simply, that non-barred galaxies could have been barred in the 
recent past. However, most numerical simulations show that, once formed, bars are 
robust structures \citep{shensellwood2004, athanassoulaetal2005, debattista2006, 
berentzen2007, villavargas2010, kraljicetal2012, athanassoulaetal2013}. 
Furthermore, at least in massive disk galaxies, bars have the same stellar 
population properties of bulges (old, metal rich, and $[\alpha/$Fe$]$-enhanced 
stellar populations; \citealt{sanchezblazquez2011}; \citealt{perez2009}) which, in 
many cases, are very different from that of the disk (see Sect.~\ref{sec:bars}). 
This result also supports (although it does not prove, see 
Sect.~\ref{sec:bars}\footnote{The stars in the bar can form in the disk long time 
ago, even if the bar have been recently formed.}) the idea that bars formed long 
ago. The longevity of bars is also suggested in studies of the bar fraction 
evolution \citep[e.g.][]{shethetal2008}, which find a similar bar fraction at 
z$\sim$0.8 to that seen at the present-day for galaxies with stellar masses 
$M_{*} \geq 10^{11}$ M$_{\odot}$. In addition, non-axisymmetric structures, such 
as nuclear spirals, can drive gaseous inflows \citep[e.g.][]{kormendyfisher2005}, 
which could dilute the differences between barred and non-barred galaxies.

\subsection{Stellar population of bars}
As the debate of the durability of bars is still open and its influence 
may be crucial for the formation of bulges, it is also important to 
study the stellar populations hosted by bars. Very few works, however, 
have dealt with this problem.

\citet{gadottidesouza2006} obtained the color and color gradients in the bar 
region of a sample of 18 barred galaxies. They interpreted the color differences 
as differences in stellar ages and conclude that younger bars were hosted by 
galaxies of later types \citep[see also][]{gadotti2008}. However, as we mentioned 
in Sect.~\ref{sec:colors}, the effects of age-metallicity degeneracy and dust extinction are 
strongly degenerate in colors and, therefore, conclusions based on only colors 
remain uncertain. \citet{perez2007} and \citet{perez2009} performed an analysis 
of the stellar population of bars in early-type galaxies using line-strength 
indices. They found that the mean bar values of SSP-equivalent age, metallicity, 
and $[\alpha/$Fe$]$, correlate with central $\sigma$ in a similar way to that of 
bulges, pointing to an intimate evolution of both components. Galaxies with high 
central $\sigma$ ($>$170 kms$^{-1}$) host bars with old stars while galaxies with 
lower central velocity dispersion show stars with a large dispersion in their 
ages.

These authors also analyzed the stellar population gradients along the 
bars and found three different behaviors: (1) bars with negative 
metallicity gradients. These bars have young/intermediate stellar 
populations (SSP-equivalent values $<$ 2 Gyr) and have amongst the lowest 
stellar velocity dispersions of the sample; (2) bars with no 
metallicity gradients. These galaxies have, however, positive age 
gradients and (3) bars with a mean old stellar population and positive 
metallicity gradients (more metal-rich at the bar ends).

The fact that bars are composed of old stellar populations does not mean that 
they formed long ago, as the bar might have formed recently out of old stars in 
the disk. One way to disentangle these two options is to compare the stellar 
populations of the disk and the bar at the same distances. In 
\citet{sanchezblazquez2011}, this comparison is made for two galaxies, finding 
that stars in the bar are older and more metal rich than those of the disk. 
Furthermore, the gradient in both parameters is much flatter in the bar. In 
general, they found that the stellar content of the bar is more similar to that 
of the bulge than to the disk. However, the sample of this study remains small 
and biased towards early-type bulges. Clearly, a larger study sampling larger 
samples of galaxies covering all morphological types is still needed.

\section{Star formation histories}
\label{sec:sfh}

The majority of stellar population studies in bulges are based on the 
comparison of the observed colors or spectral properties with the 
predictions of SSP models. While the SSP assumption may not be a bad 
approximation for massive elliptical galaxies, it is most likely not a 
good one for spirals, which are believed to have a more extended star 
formation history \citep{kennicutt1993, jamesetal2008}.

In cases were the star formation history has been more complicated than 
just a single burst, the interpretation of the results based
on analyses of single-stellar population is difficult.
Bulges with an intermediate 
SSP-equivalent age could have formed all their stars at intermediate 
epochs or almost all at very early times and a small fraction at recent 
epochs.

\begin{figure}
\centering
\includegraphics[scale=.5]{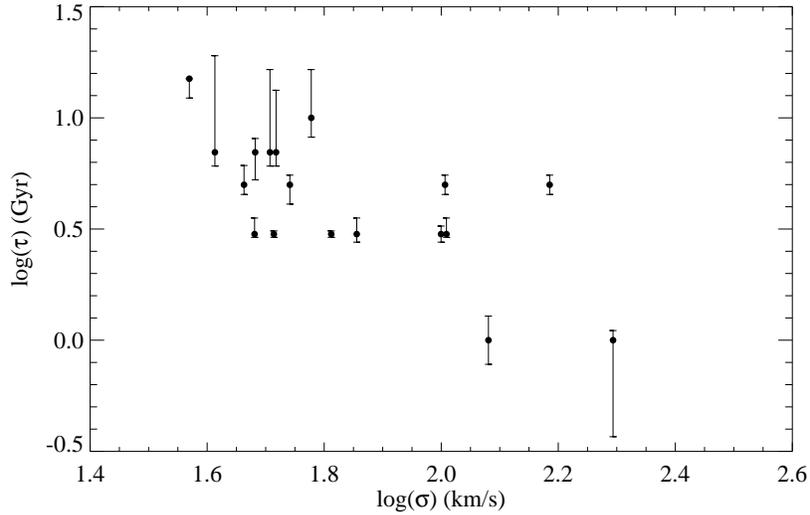}
\caption{Central aperture values for the $e$-folding time-scale $\tau$ 
against central velocity dispersion $\sigma$ (both in units of decimal 
logarithm), in an exponentially declining star formation scenario; the 
$\tau$ values are obtained selecting amongst models with age=10~Gyr.
Figure from \citet{gandaetal2007}.
\label{ganda_tau}}
\end{figure}

To avoid these difficulties, some authors have tried to analyse their data 
assuming more realistic star formation histories. There are several examples in 
the literature where the stellar population properties have been derived using a 
parametric approach \citep{gandaetal2007, macarthuretal2004, kauffmann2003} where 
a predefined shape for the star formation history and chemical enrichment is 
assumed. In this case, some parameters are fixed while others are fitted by 
comparing the observations with the predictions of the models. 
\citet{gandaetal2007} compared three line-strength indices with the predictions 
based on two burst, and on exponentially declining star formation histories. They 
obtained, respectively, the age, metallicity and mass fraction of the youngest 
burst, and the e-folding ($\tau$) time in the case of the exponentially declining 
star formation history. In the first case, they concluded that degeneracies in 
the parameter space prevented them from extracting useful conclusions. In the 
second case, they found that bulges with a larger central $\sigma$ showed shorter 
e-folding times, more consistent with an instantaneous burst scenario. 
Low-$\sigma$ galaxies have larger $\tau$, indicating a more extended star 
formation history (see Fig.~\ref{ganda_tau}).

The problem with this approach is that the results depend strongly on 
the priors -- i.e., the wrong answer can be obtained with the wrong 
assumptions about the star formation history. Furthermore, the use of 
only a few line-strength indices makes it difficult to break the 
existing degeneracies in the parameter space, such as the age of the 
burst versus its strength, or the $\tau$-metallicity degeneracy.

\begin{figure}
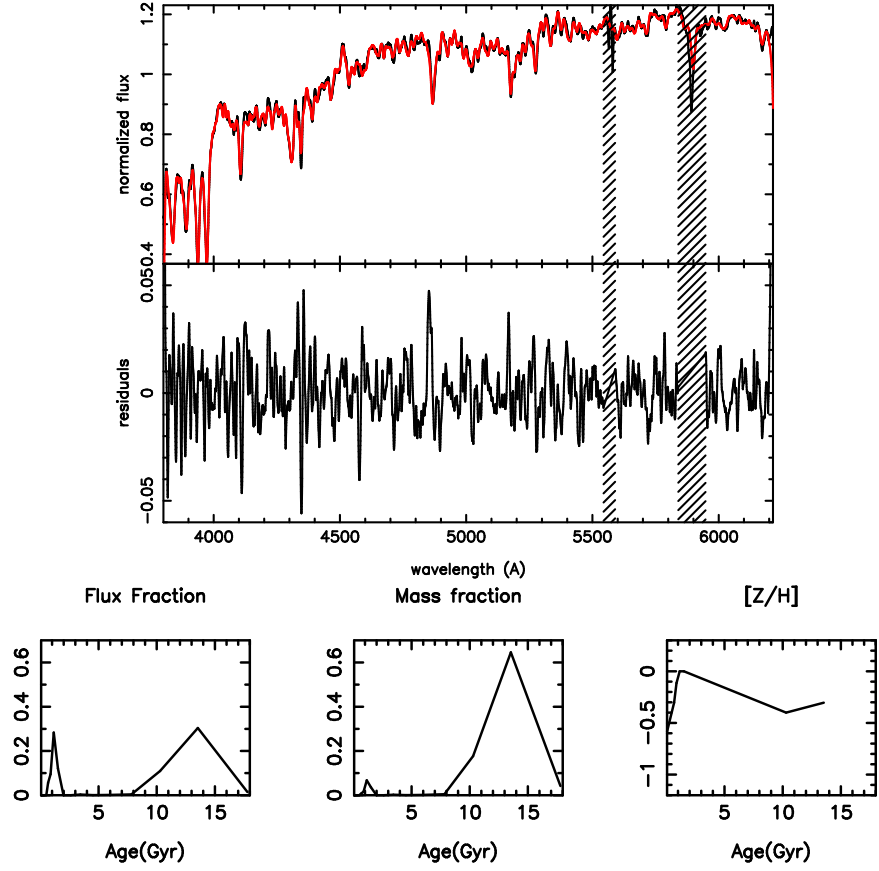

\centering
\includegraphics[width=0.66\hsize,angle=-90]{fit_integrated.ps}
\includegraphics[width=0.32\hsize,angle=-90]{figura_paper.ps}
\caption{Top panel: Example of the fit (red line) obtained with {\tt 
STECKMAP} for the integrated spectrum of NGC~628 (black line). The 
residuals from the fit (observed-fitted) are also showed.
Hashed regions indicate those zones that were 
masked during the fit. Bottom panels: Mass, flux fractions and the 
age-metallicity relation derived with {\tt 
STECKMAP} for the integrated spectrum of NGC~628. From 
\citet{sanchezblazquez2014a}.
\label{fig:psb2014}}
\end{figure}

The availability of high-quality stellar libraries and associated stellar 
population models \citep[e.g.][]{sanchezblazquez2006, vazdekisetal2010, 
bruzualycharlot2003, conroyvandokkum2012} that predict, not only individual 
absorption line features but the whole spectral energy distribution, has allowed 
the development of new techniques that, fitting the whole spectrum, are able to 
obtain, not only SSP-equivalent ages and metallicities, but a more, in principle 
realistic, star formation history. Furthermore, by considering the information 
provided by the entire spectrum, the age and the metallicity are more easily 
separated \citep{sanchezblazquez2011}. These techniques are non-parametric -- 
i.e., no predefined shape for the star formation history is assumed. Codes that 
have been used to study the stellar population properties of galaxies include 
MOPED \citep{heavens2000}; VESPA \citep{tojeiroetal2007}; STECKMAP 
\citep{ocvirk2006a}; STARLIGHT \citep{cidfernandesetal2005}; SEDFIT 
\citep{walcher2006} and ULySS \citep{koleva2009}. Using these new tools, one can 
fit an observed spectrum in terms of a model built by a linear combination of a 
number of SSPs with different ages and metallicities. The kinematics can be 
calculated at the same time by convolving the model with a Gaussian line-of-sight 
velocity distribution. In some cases, dust can be modeled assuming a reddening 
law. Figure~\ref{fig:psb2014} shows an example of a fit to the integrated 
spectrum of NGC~628 using the code {\tt STECKMAP} \citep{ocvirk2006a, 
ocvirk2006b}, together with the derived flux and mass fractions for stars of 
different ages and the age-metallicity relation \citep{sanchezblazquez2014a}.

\begin{figure}
\centering
\includegraphics[width=\hsize]{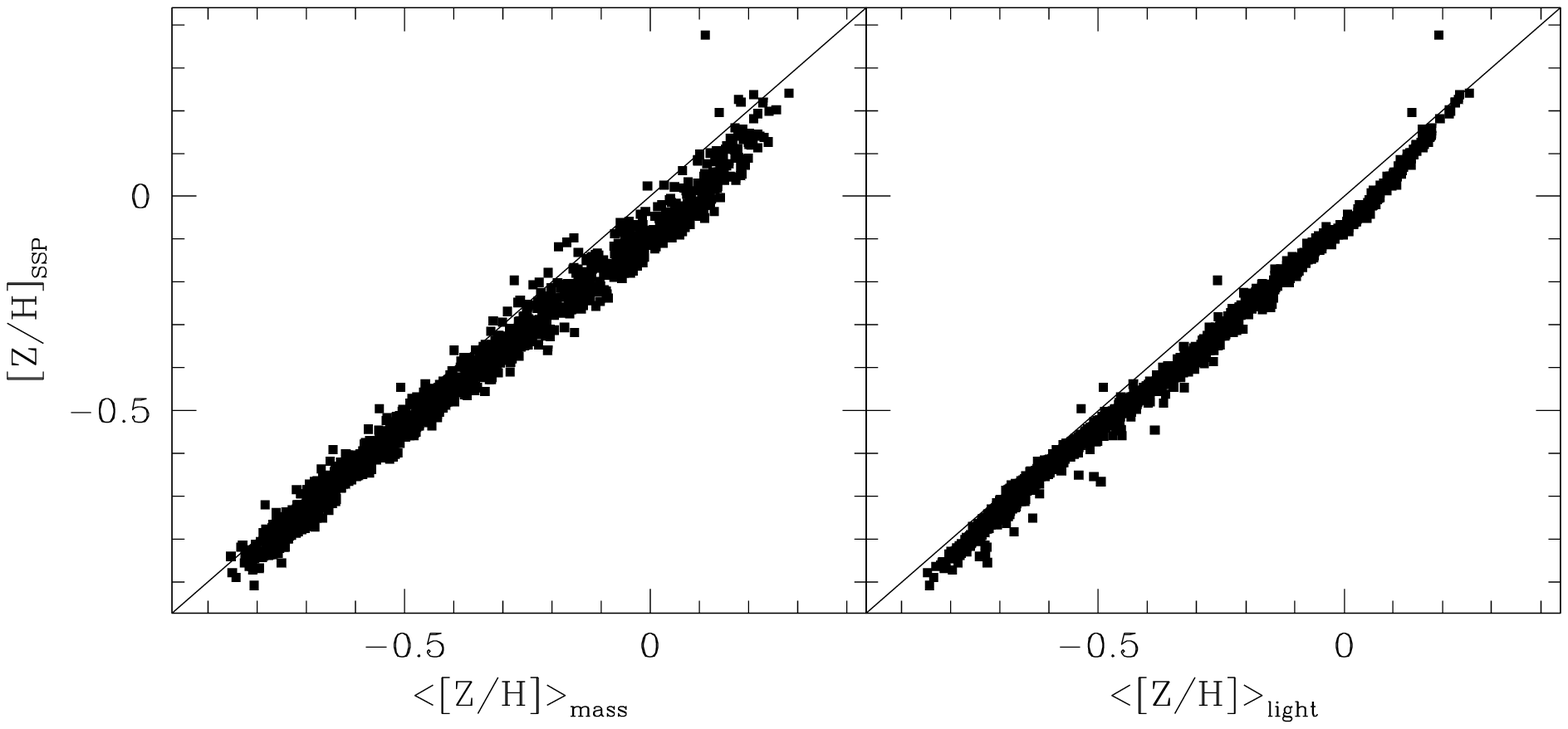}
\includegraphics[width=\hsize]{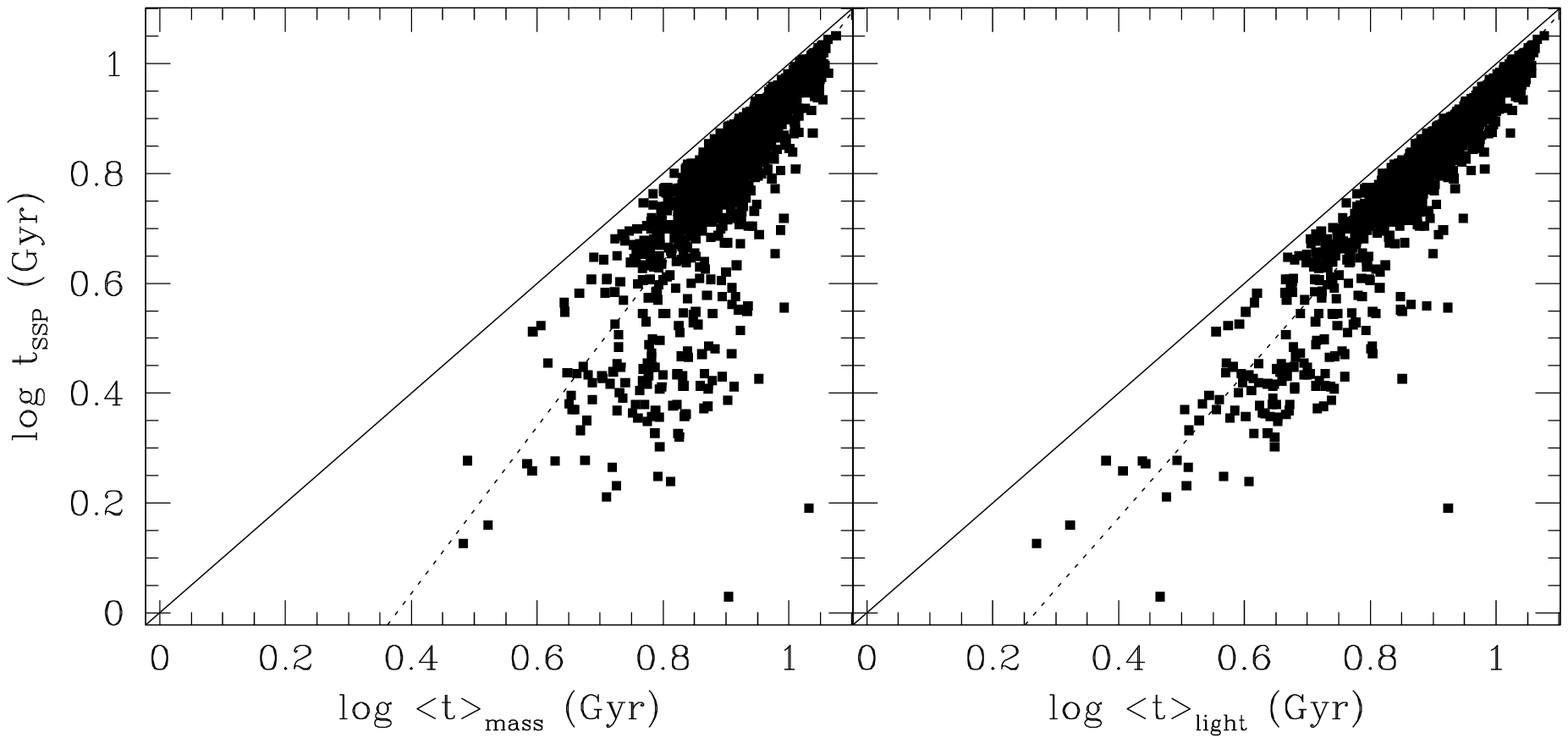}
\caption{\label{tragersomerville} Comparison of the 
SSP-equivalent age and metallicity as a function of mass-weighted 
(left-hand panels) and V-band light-weighted values (right-hand panels) 
for models of early-type galaxies drawn from 20 realizations of a Coma 
cluster-size halo. Figure from \citet{tragersomerville2009}.}
\end{figure}

The problem of inverting a spectrum to derive detailed star formation and 
chemical enrichment histories is ill-conditioned -- i.e., small fluctuations in 
the data can produce strong variations in the final solution -- and different 
codes try to overcome these issue with different methods. The accuracy of the 
recovered star formation history depends critically on the signal-to-noise of the 
input spectra and, depending on this value, one can recover more or less 
different stellar populations described by an age and a metallicity. However, 
with spectra of enough quality, the methods have shown that they can recover 
reliably both the age distribution and the age-metallicity relation \citep[see, 
e.g.][]{cidfernandesetal2005, ocvirk2006a, sanchezblazquez2011}.

Robust quantities, even when derived from a spectrum with low 
signal-to-noise, are the mean values of age and metallicity 
\citep{cidfernandesetal2005} that are usually  
weighted with the light (LW) or the mass (MW) of the stars.  These are obtained as:
\begin{eqnarray}
<A>_{{\rm LW}} = \frac{\sum_{i}^N A_i \times {\rm flux}_i}{ \sum_i {\rm flux}_i}\\
<A>_{{\rm MW}} = \frac{\sum_{i}^N A_i \times {\rm mass}_i}{ \sum_i {\rm mass}_i},
\end{eqnarray}
where $A$ represents the physical parameter  (age or 
metallicity) and mass$_i$ and flux$_i$ are, respectively, the 
reconstructed mass and flux contributions of the stars in the $i$-th age 
bin, as returned by the code. When present, young stars are very 
luminous in the optical and, therefore, contribute more 
to the light-weighted values. This means that the light-weighted values 
of age will be biased towards the youngest stellar components. The 
mass-weighted values will be less biased but they are also more 
uncertain, as the contribution to mass by low-mass faint stars can be 
very important.

It is interesting to make a comparison between the SSP-equivalent parameters and 
the averaged ones obtained from the full star formation history. This comparison 
was made by \citet{tragersomerville2009}. In their work, the authors derived 
stellar population parameters from synthetic spectra generated by a hierarchical 
galaxy formation model. Figure~\ref{tragersomerville} shows the comparison of the 
SSP-equivalent ages and metallicities with the mean values weighted with both 
light (in the V-band) and mass. As can be seen, the SSP-equivalent ages are 
always lower than both the luminosity- and mass-weighted averages. In particular, 
SSP-equivalent ages reflect more closely the age of the last episode of star 
formation while luminosity-weighted means, although still biased towards the ages 
of the youngest components, are closer to the unweighted mean. On the other hand, 
SSP-equivalent metallicities and abundance ratios are less severely biased.

\begin{figure}
\centering
\includegraphics[scale=.50]{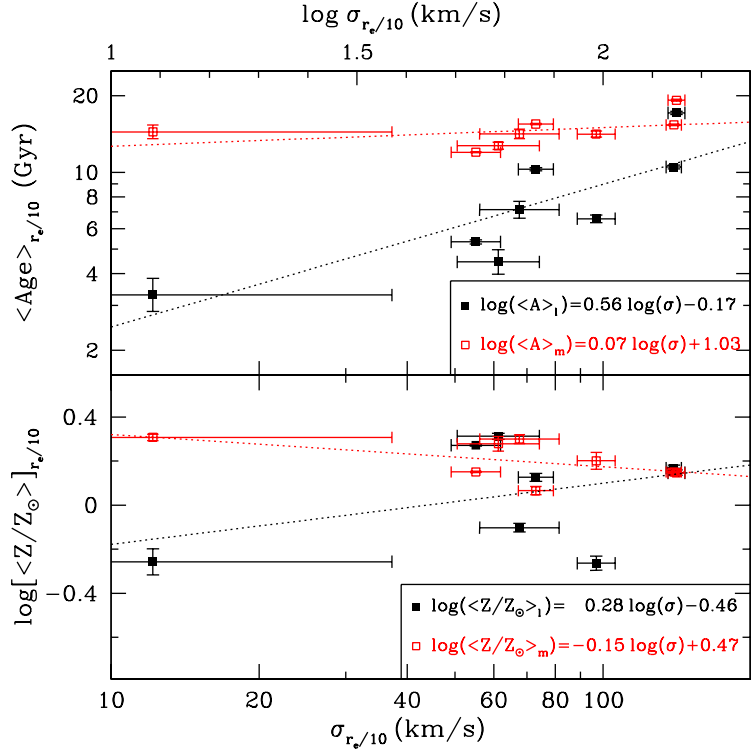}
\caption{Average age and metallicity as a function of velocity 
dispersion for a sample of face-on bulges. Black solid squares represent 
light-weighted values while red symbols are mass-weighted. The black and 
red dotted lines are linear regressions to the light- and mass-weighted 
data. From \citet{macarthur2009}.
\label{ma2009}}
\end{figure}

Some studies have used the non-parametric techniques to analyze the star 
formation histories of galaxy bulges \citep{macarthur2009, sanchezblazquez2011, 
gonzalezdelgado2014}. The first conclusion from these studies is that the SSP is 
a very bad approximation for the star formation history of these objects. 
Furthermore, \citet{macarthur2009} and \citet{sanchezblazquez2011} found that in 
a mass-weighted context, all bulges in their sample were predominantly composed 
of old stars, independently of their central velocity dispersion. In fact, the 
previously reported trends of age with central velocity dispersion disappear when 
mass-weighted values of age are used instead of light weighted ones. This can be 
seen in Fig.~\ref{ma2009}, from \citet{macarthur2009}.

The result that bulges are dominated, in mass, by old stars applies to 
all types of bulges (early- and late-type, showing different S\'ersic 
index, and with and without bars). However, this type of analysis has 
been performed on a very low number of galaxies. It would be desirable 
to extend this work to a complete sample of bulges covering a range of 
masses and morphological types. Ideally, one would like to quantify the 
mass contribution of the young component and correlate this with other 
properties of the bulges, such as the mass, the environment, the 
presence of bars, and the spiral arm morphology. This would allow one to 
study the importance of secular versus external mechanisms in building 
up the central bulges.

Recent works deriving the star formation history for large samples of 
spiral galaxies are those performed on data from the CALIFA survey 
\citep{perez2013, gonzalezdelgado2014,  
sanchezblazquez2014}. None of these studies has yet 
especially investigated the population in the bulge, but 
\citet{perez2013} analyze the mass assembling history of the central 
parts of galaxies compared with the rest of the galaxy for galaxies of 
all morphological types binned in mass. They found that galaxies with 
stellar mass $M_{*} > 5 \times 10^{10}$ M$_{\odot}$ have grown quickly 
their inner part, 5-9 Gyr ago, while lower mass galaxies formed their 
stars more slowly.

One caveat to all these studies performing full spectral fitting is that 
they use stellar population models with chemical abundance ratios scaled 
to solar \citep[e.g.][]{bruzualycharlot2003, vazdekisetal2010}. This 
implies that the models are tuned to the specific chemistry and star 
formation history of our MW. This is because the empirical spectral 
libraries are limited to those stars in the solar neighborhood.

While predictions of the strengths of Lick/IDS indices with variable abundance 
ratios have been made for SSPs of different ages and metallicities 
\citep{trageretal2000, thomas2003, proctoretal2004, tantalochiosi2004, 
Leeworthey2005, annibalietal2007, schiavon2007} using a semi-empirical approach, 
the calculation of the entire spectral energy distribution is more challenging. 
However, in the last few years, full spectrum fitting models have been extended 
to include variation in the elemental abundance patterns, using either 
theoretical stellar libraries or semi-empirical approaches \citep{coehloetal2007, 
walcheretal2009, conroyvandokkum2012}. The first studies using full spectral 
fitting to derive chemical abundance ratios of different elements are starting to 
appear in the literature, all using samples of early-type galaxies 
\citep{conroyetal2014, walcheretal2009}. This is, however, a challenging task, 
due to the large number of parameters to fit and the possible degeneracies 
between them.

\section{Spatially resolved  stellar populations in bulges}
\label{sec:gradients}

Most stellar population studies in bulges have been done using the 
integrated properties inside a certain aperture. In case of 
spectroscopic studies, this aperture commonly encloses just the very 
central parts. However, if we want to have a full understanding of bulge 
formation, it is necessary to gain knowledge of the variations of the stellar 
populations with radius. These variations are intimately connected with 
the dynamical processes that led to the formation of these structures, 
the degree of dissipation, and the possible re-arrangement of material.

Mergers with gas dissipation or monolithic collapse scenarios predict steep 
metallicity gradients \citep{eggen1962, larson1974,arimotoyoshii1987} and strong 
gradients in $[\alpha/$Fe$]$ \citep{ferrerassilk2002}. The predictions for 
secularly formed bulges are more complicated. As they formed from redistribution 
of disk stars, the final metallicity gradient will depend on the original 
gradient in the disk and the scale-length of the final bulge and also on the disk 
heating \citep{moorthyholtzman2006}. However, lower metallicity gradients are 
expected compared to those of the first scenario. Observationally, the Galactic 
bulge, which manifests many characteristics of a peanut-shaped bulge, has a clear 
vertical metallicity gradient, such that the more metal-rich part of the 
metallicity distribution thins out towards high latitudes \citep{minnitietal1995, 
zoccalietal2008, gonzalezetal2011}. This result has long been taken as a 
signature for a classical bulge in the MW. However, recent results have shown 
that the stars that have been scattered furthest from the disk are the oldest 
stars and, consequently formed from the least metal-enriched fuel 
\citep{freeman2008, martinezvalpuesta2013}. The buckling process may hence 
establish a negative minor-axis metallicity gradient (which is observed in the MW 
and NGC~4565 \citep{proctor2000}.

\begin{figure}
\centering
\includegraphics[trim=0 0 0 40,clip,width=0.78\hsize,angle=90]{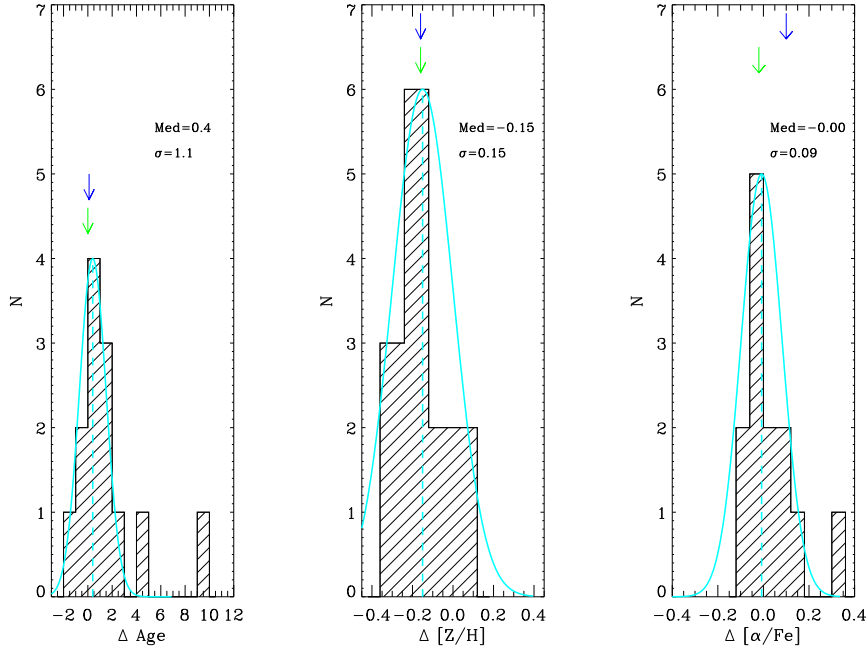}
\caption{Distribution of the gradients of age (left), metallicity 
(central), and $[\alpha/$Fe$]$ enhancement (right) for a sample of 
galaxies. The dashed line represents the median of the distribution and its 
values is also reported. The solid line represents a Gaussian centered on the 
median value of the distribution. Their $\sigma$ is approximated by the value 
containing the 68\% of objects of the distribution and is noted in the inset 
label. The green and blue arrows show the average gradient found for 
early-type galaxies and bulges by \citet{mehlert2003} and 
\citet{jablonkaetal2007}, respectively (figure from 
\citet{morellietal2008}).
\label{fig12_morelli}}
\end{figure}

Several articles have studied the variation of the spectral features with radius 
in bulges \citep{moorthyholtzman2006, jablonkaetal2007,morellietal2008, 
perezsanchezblazquez2011, macarthur2009, sanchezblazquez2011, gandaetal2007} and 
compared them with stellar population models to obtain either SSP-equivalent 
parameters or mean values, based on the recovery of the star formation history. 
These studies find that most bulges have SSP-equivalent or luminosity-weighted 
negative gradients in the metallicity, almost no gradients in age, and slightly 
positive or null $[\alpha/$Fe$]$ gradients. Metallicity gradients in the bulge 
regions are generally steeper than those in the disk region 
\citep{moorthyholtzman2006, sanchezblazquez2011}. Figure~\ref{fig12_morelli} 
shows the distribution of the SSP-equivalent gradients for a sample of bulges 
taken from the work of \citet{morellietal2008}.

These values are similar to the ones found in elliptical galaxies, 
although the quantitative comparison is not that clear. 
\citet{jablonkaetal2007} and \citet{morellietal2008} do not find any 
difference in the magnitude of the bulge gradients and those of elliptical 
galaxies, while \citet{williamsetal2012}, on the other hand, found that 
the gradients in boxy bulges are shallower than those in elliptical 
galaxies at a given $\sigma$.

Several studies have also looked for correlations between the gradients and other 
properties of the galaxies, such as the central $\sigma$, the luminosity or the 
mass, to check if they are related with the potential well of the galaxy, as 
the central values are. \citet{goudfrooijetal1999} and \citet{proctor2000} found 
that gradients were correlated with luminosity and central $\sigma$, 
respectively, although from very small samples, while \citet{jablonkaetal2007} 
found no such correlation. However, there seems to be a trend for which small 
bulges have lower gradients \citep[see also][]{moorthyholtzman2006, 
gonzalezdelgado2014}. This, in principle, could be attributed to the fact that 
secularly formed bulges are more common in low mass galaxies.

The possible differences between the gradients of bulges with and without bars 
have also been explored in a few works, and none of them find any significant one 
\citep{moorthyholtzman2006,jablonkaetal2007, perezsanchezblazquez2011}. However, 
\citet{moorthyholtzman2006} reported that when a positive age gradient was 
present, it was always in barred galaxies, which could indicate that these 
objects have more extended star formation in their centers due to bar-driven 
inflow of gas. This result agrees with that of \citet{gadottidosanjos2001}, who 
found a greater prevalence of null or positive color gradients in barred galaxies 
than in non-barred galaxies, which they interpret as an evidence for gradients 
being erased by bar-driven mixing. Nonetheless, this has not been confirmed in 
other studies \citep{jablonkaetal2007}. The reason for the discrepancies could 
be, once again, the orientation of the samples. The majority of authors agree 
that positive age gradients are normally the consequence of the the presence of 
central disks or nuclear rings with recent star formation 
\citep[e.g.][]{morellietal2008}. As central disks and rings are likely more 
common in barred galaxies, this can explain the differences between barred and 
non-barred galaxies found by \citet{moorthyholtzman2006} and also the lack of 
differences reported in the edge-on sample of \citet{jablonkaetal2007}, as these 
flattened central structures will not contribute to the observed light of the 
bulge in these orientations.

This interpretation of the age gradients as the consequence of the presence of 
central younger structure is supported by the fact that the mass-weighted age 
gradients are, in the majority of cases, much flatter than the 
luminosity-weighted or the SSP-equivalent ones, indicating the the majority of 
the stars in the bulge are old and share a common age, while a small fraction of 
stars concentrated in central structures are causing the observed radial trends 
\citep{macarthur2009, 
sanchezblazquez2011,sanchezblazquez2014,gonzalezdelgado2014}.

Interestingly, \citet{jablonkaetal2007} find that the line-strength 
indices at 1 $r_{{\rm eff}}$\footnote{The radius that contains half of the total 
luminosity of the bulge.} were very similar for all the galaxies, 
independent of their mass or morphological type, and the different 
gradients come from the differences in their central indices.

An issue in measuring the gradients of age, metallicity and 
$[\alpha/$Fe$]$ in bulges could be the contamination of their stellar 
population by the light coming from the underlying disk stellar 
component. This effect is not important in the galaxy center but it can 
have an enormous impact in the stellar population estimates of the 
external parts where the contribution from the disk to the total light 
is  more important. Different authors have tried to quantify one 
way or the other this contamination from the disk to the bulge light 
 \citep[see][]{jablonkaetal1996, moorthyholtzman2006, morellietal2008} and it 
does not seem to be very important, but the number of tests is small and disk 
contamination is still an issue in the measurement of gradients. Studies of 
edge-on galaxies \citep{jablonkaetal2007} do not have this problem although they 
have the drawback of being blind to flatter components in the center of the 
galaxy.

\section{The bulge-disk connection}
\label{sec:connection}

The age distribution of disks is of obvious importance for constraining scenarios 
of disk-bulge formation. Correlations between the disk and bulge colors have been 
found by several authors \citep{peletierbalcells1996, dejong1996, belldejong2000, 
carolloetal2001, gadottidosanjos2001} which stands for both early and late type 
spirals. The similarity in color between inner disk and bulge has been 
interpreted as implying similar ages and metallicities for these two components 
and an implicit evolutionary connection \citep{dejong1996, peletierbalcells1996}. 
However, using only colors one cannot directly transform the correlation in 
colors into correlations of age and/or metallicity.

A correlation between the line-strength index $[$MgFe$]'$ at one disk 
scale length and in the center was also found by 
\citet{moorthyholtzman2006}.  They interpret this as a correlation 
between the metallicity of bulge and disk as this index is more 
sensitive to variations of metallicity than age and it is almost 
insensitive to variations of $[\alpha/$Fe$]$.

\begin{figure}
\centering
\includegraphics[scale=.5,angle=-90]{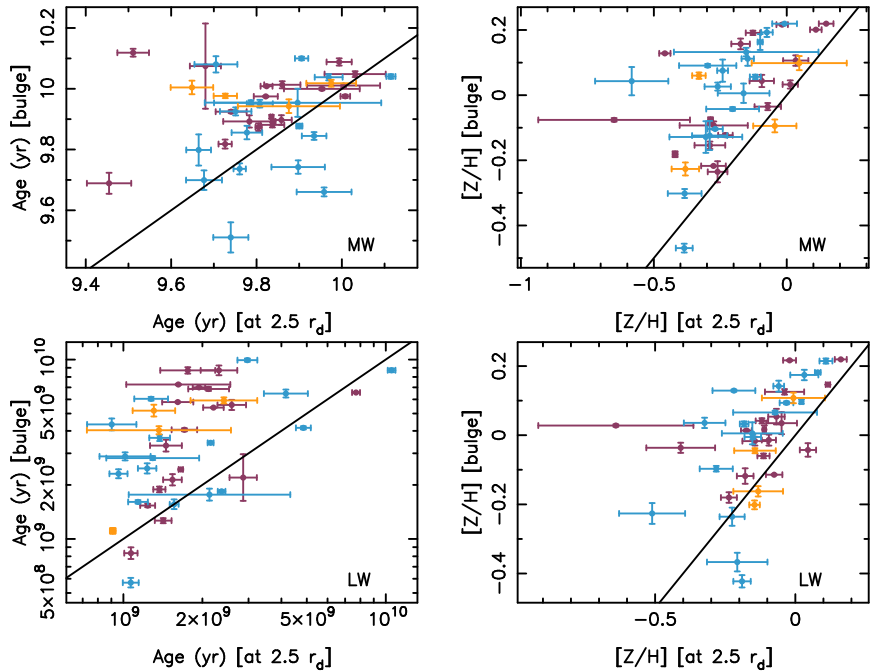}
\caption{Comparison of the average ages (left panels) and metallicity 
(right panels) weighted by both mass (top panels) and light (bottom 
panels) measured in the center and at 2.5 scale-lengths of the disk. 
Different colors indicate if the galaxy is barred (red), non-barred (blue) 
or weakly barred (orange). The solid line represents the 1:1 relation. 
For more details see \citet{sanchezblazquez2014}.
\label{fig:bulge_disk}}
\end{figure}

More recently \citet{sanchezblazquez2014} performed a full spectral fitting 
analysis of a sample of 62 nearly face-on spiral galaxies observed as a part of 
the IFS survey CALIFA \citep{sanchezcalifa} and showed that the slope of the 
relation between the luminosity-weighted age and metallicity and central velocity 
dispersion is similar for the central parts of the bulges and for the disk at 
$\sim$2.4 scale-lengths, but the bulges show higher luminosity-weighted ages and 
metallicities. Figure~\ref{fig:bulge_disk} shows the comparison of the mean 
values of age and metallicity (both weighted with mass and light) in the center 
and at 2.5 scale-lengths of the disk. A Spearman rank order tests reflect that, 
while the metallicities of both components are correlated, the same does not 
happen with the ages.

Note that the correlation exists for all early and late-type spirals, 
spherical and pseudobulges, contrary to the connection between the 
radial sizes of bulges and disks, that only exists for pseudobulges 
\citep{fisherdrory2008}. Therefore either all bulges formed secularly 
and some had their bar destroyed or other physical processes are 
responsible for this correlation. Several authors have shown that the 
correlation between the disk and bulge sizes can exist in major and 
minor merger remnants. While minor mergers tend to preserve the original 
bulge-disk coupling of the main progenitor, major mergers are capable of 
rebuilding a bulge-disk coupling from the remnants after having destroyed 
the original structures of the progenitors \citep[][and 
references therein]{querejeta2014}.

On the other hand, \citet{morellietal2012} rule out significant 
interplay between the bulge and disk components due to the similarity in 
the stellar population properties of bulges hosted in galaxies with very 
different disks (high- and low-surface brightness).

\section{Summary}
\label{sec:summary}

The study of stellar populations can provide the critical test needed to 
understand the basic mechanism driving bulge formation. However, a clear 
picture about the stellar population in bulges and their possible 
correlations with other parameters is still not obvious. We have collated 
the main results obtained from the study of colors and spectral 
characteristics and the comparison with stellar population models and
can thus summarize them: 

\begin{itemize}
\item Bulges show a wide range of SSP-equivalent ages and metallicities. 
There is a trend for which more massive bulges have, on average, older 
stellar populations and higher values of metallicity and 
$[\alpha$/Fe$]$.
\item Bulges with high S\'ersic index $n$ tend to be old and have high 
[$\alpha$/Fe] but it is not clear if this trend is due to the existing 
trend between mass and $n$.
\item There are not strong differences in the stellar populations of 
bulges and elliptical galaxies at the same mass, but the details are of 
this comparison are still not clear and conclusions differ between 
studies.
\item Bulges are, in many cases, composite systems, with disky, 
boxy/peanut and classical bulges coexisting in the same galaxy. They 
usually host different stellar populations. Young stars, when present, 
are located in central disks and rings. These young components are not 
seen in edge-on samples which has lead studies differing in the 
orientation of the sample to obtain different results.
\item In general terms, there are not very clear differences in the 
stellar population properties of bulges with and without bars, neither 
in the central values nor in the variations with radius. If there are 
differences, those are only present in massive galaxies, and they show 
up as an excess of young populations when compared to unbarred galaxies.
\item When the mass-weighted mean values of age are considered, all 
bulges, independent of their mass, seem to be dominated by old stars. 
The trends between age, metallicity, and mass become much flatter and 
almost non-existent. This is true for bulges with a variety of structural 
parameters, such as different S\'ersic indices or surface brightness 
profiles.
\item Bulges show mild, negative, metallicity gradients and almost 
null $[\alpha/$Fe$]$ and age gradients. The distribution of the slopes 
is similar to that found in elliptical galaxies. At present, there is 
no agreement about the possible correlation of these 
gradients and other parameters.
\item There is a correlation between the metallicity of the bulge and 
the disk but not between the ages.
\end{itemize}

\section{Future prospects}
\label{sec:future} 

Although considerable progress has been made in the field, there is 
still much to be done, in order to understand the star formation 
histories and chemical evolution in the bulges of spirals and S0 
galaxies. The first thing we have to deal with is the fact that we 
cannot make a clean distinction between classical, disky, and boxy/peanut 
bulges, as many of them coexist in the same galaxy. We need to quantify 
the preponderance of each component and correlate this with other 
properties of the galaxies, like the mass, the environment, the presence 
of bars, type of spiral arms, etc. We need also to understand the 
physical mechanism that formed each components. For example, numerical 
simulations have shown that bulges with structural characteristics of 
pseudobulges can be formed, not only secularly, but also quickly, at 
high redshift, via a combination of non-axisymmetric disk instabilities 
and tidal interactions or mergers.

\begin{figure}
\centering
\includegraphics[width=0.49\hsize]{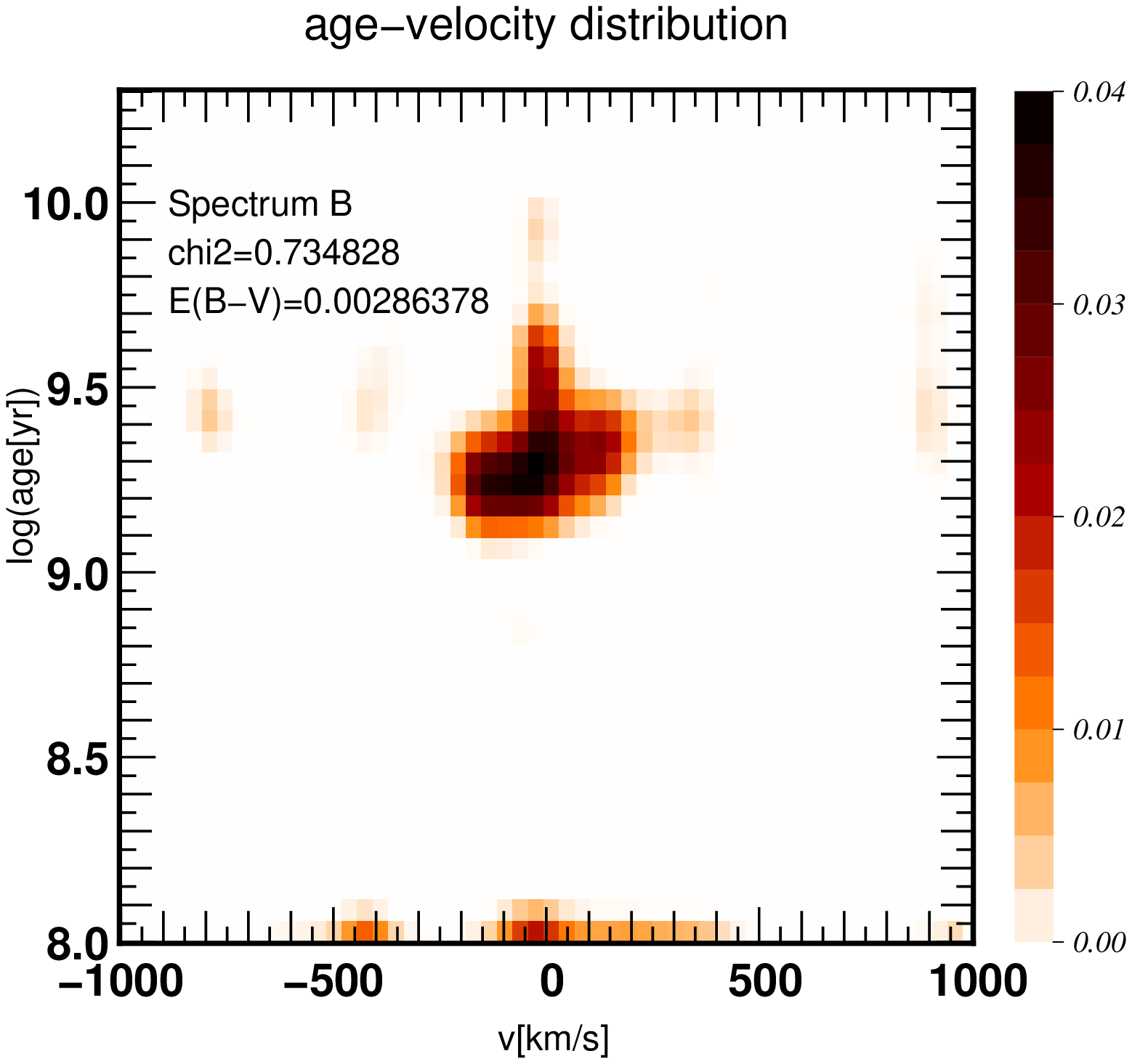}
\includegraphics[width=0.49\hsize]{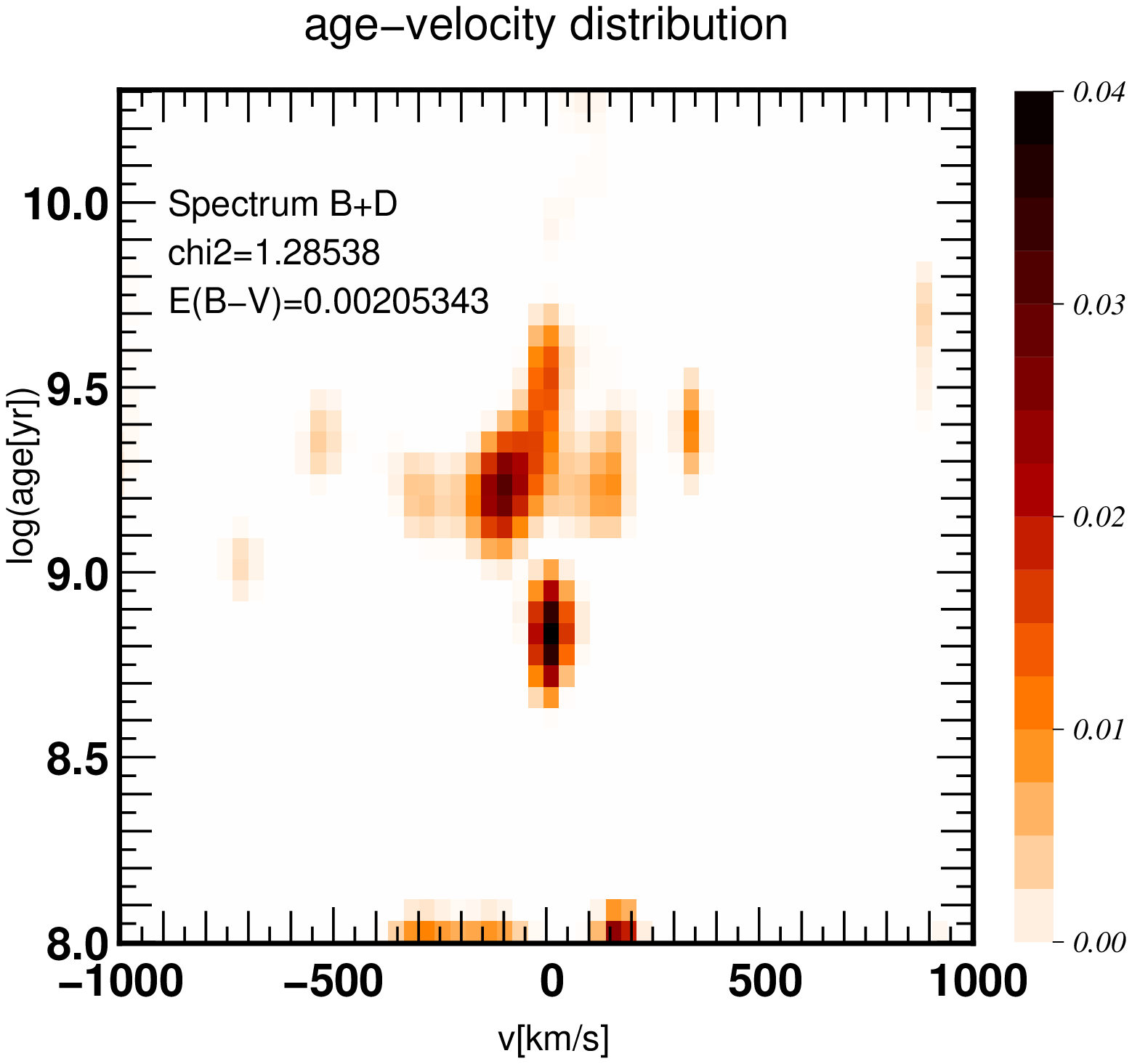}
\caption{\label{fig:ocvirk2008} Age-velocity map reconstruction for the 
center of the bulge spectrum of NGC~4030 (left) and for the bulge spectrum 3 
arcsec from the center (right panel), where a disk component appears on 
top of the dynamically hot component.  From \citet{ocvirk2008}.}
\end{figure}

Most advances, therefore, will come from the use of new techniques to derive star 
formation histories that allow us to distinguish different episodes of star 
formation. The future is promising: \citet{ocvirk2008} and \citet{coccato2011} 
show the feasibility of separating different components, not only in terms of 
stellar populations, but also kinematically. This is illustrated in 
Fig.~\ref{fig:ocvirk2008}, from \cite{ocvirk2008}, where the age-velocity 
distribution for the bulge is shown in two positions (in the center and outside 
the bulge dominated region of the galaxy). In this work, the authors model the 
observed spectrum as a sum of 40 components with different ages but, contrary to 
the traditional assumption that all components share the same line-of-sight 
velocity distribution (LOSVD), each component was allowed to have its own 
non-parametric (not necessarily Gaussian) LOSVD. These authors reconstructed the 
age-velocity distribution of two bulge regions of the Sbc galaxy NGC~4030 and were 
able to separate two components, one with a relatively young stellar population 
($\sim$2 Gyr) and $\sigma\sim 100$~km/s and an even younger ($\sim$ 500 Myr), and 
kinematically colder ($\sigma\sim 30$~km/s). Its location (outside the 
bulge-dominated region of the galaxy), kinematics, and relatively young age 
suggest that this is a young internal disk. The null mean velocity of the 
structure is expected for the minor axis of a stellar disk. This study 
demonstrates the feasibility of separating and measuring the age and kinematics 
of superimposed galactic components from an integrated light spectrum. This type 
of analysis will be very useful in characterizing the properties of the different 
components in composite bulges and isolating the properties of their stellar 
populations.

Furthermore, we have mentioned throughout the chapter 
several results that need to be  confirmed with larger samples.   
Spectroscopic studies of a large sample of bulges with different 
characteristics are clearly needed to advance in our understanding of 
bulge stellar populations. Ideally this sample will be observed with an 
integral field unit --to be able to separate morphologically the 
different subcomponents -- and it will contain galaxies in different 
environments. Indeed, the influence of environment has not been studied 
thoroughly. \citet{morellietal2008} analyzed a sample of bulges in 
the Coma cluster but they did not compare their results with 
bulges in other environments. \citet{peletieretal1999} did not find any 
difference in the colors of galaxies in different environments and 
concluded that this parameter does not have a strong influence in 
shaping the star formation in bulges. Nonetheless, a more comprehensive 
study remains to be done. A detailed study of the properties of bulges 
with environment might help us to distinguish between external and 
internal processes for the formation of bulges.

We expect that instruments that are already operating, such as MUSE 
\citep{baconetal2010}, with superb spatial resolution, will help us to resolve 
central components, such as nuclear disks or bars. These data will allow the 
coupling of stellar population and kinematical properties with the morphological 
characteristics of the galaxies, improving our understanding of bulges in a way 
it has not been possible before.

\begin{acknowledgement}

I would like to thank Brad Gibson, Jairo Mendez-Abreu, Pierre Ocvirk, Jes\'us 
Falc\'on-Barroso, Adriana de Lorenzo-C\'aceres and also the anonymous referee for 
suggestions that improved this manuscript. I would also like to thank all the 
authors who have kindly provided me with figures from their publications.

\end{acknowledgement}

%
\bibliographystyle{mn2e}
\bibliography{references}
\clearpage
\end{document}